\documentclass[twocolumn,aps,pra,showpacs]{revtex4}
\usepackage{graphicx}
\usepackage{amsmath}
\newcommand{\grscale}{0.46}
\begin{document}
%%%%%%%%%%%%%%%%%%%%%%%%%%%%%%%%%%%%%%%%%%%%%%%%%%%%%%%%%%%

\title{Combined quantum state preparation and laser cooling of a continuous beam of cold atoms}

\author{Gianni Di Domenico}
\email{gianni.didomenico@unine.ch}
\author{Laurent Devenoges}
\author{Claire Dumas}
\author{Pierre Thomann}

\affiliation{Laboratoire Temps-Fr\'equence, Universit\'e de Neuch\^atel, \\
Avenue de Bellevaux 51, CH-2009 Neuch\^atel, Switzerland}

\date{7 November 2010}

\begin{abstract}
We use two-laser optical pumping on a continuous atomic fountain in order to prepare cold cesium atoms in the same quantum ground state. A first laser excites the $F=4$ ground state to pump the atoms toward $F=3$ while a second $\pi$-polarized laser excites the $F=3\rightarrow F'=3$ transition of the D2 line to produce Zeeman pumping toward $m=0$. To avoid trap states, we implement the first laser in a 2D optical lattice geometry, thereby creating polarization gradients. This configuration has the advantage of simultaneously producing Sisyphus cooling when the optical lattice laser is tuned between the $F=4\rightarrow F'=4$ and $F=4\rightarrow F'=5$ transitions of the D2 line, which is important to remove the heat produced by optical pumping. Detuning the frequency of the second $\pi$-polarized laser reveals the action of a new mechanism improving both laser cooling and state preparation efficiency. A physical interpretation of this mechanism is discussed.
\end{abstract}



\maketitle
%%%%%%%%%%%%%%%%%%%%%%%%%%%%%%%%%%%%%%%%%%%%%%%%%%%%%%%%%%%%%%%%%%%%%%%%%%%%%
\section{Introduction}
\label{section1}
%%%%%%%%%%%%%%%%%%%%%%%%%%%%%%%%%%%%%%%%%%%%%%%%%%%%%%%%%%%%%%%%%%%%%%%%%%%%%

Quantum state preparation, i.e. preparation of an ensemble of quantum systems in a given quantum state, is useful in many physics experiments ranging from quantum information science to quantum metrology, not to mention fundamental physics experiments. For example in cesium atomic clocks, one needs to create a population inversion on the clock transition ($F=3,m=0\rightarrow F=4,m=0$) before being able to probe the transition probability with Ramsey microwave spectroscopy. Indeed, when the microwave is tuned to the clock transition, the population of the $F=4$ state increases by $\Delta P = P_{3,0}-P_{4,0}$ where $P_{F,m}$ denotes the relative population of state $\left|F,m\right\rangle$. Therefore, any increase of $P_{3,0}$ will result in a corresponding increase of the clock resonance signal and as a consequence of the clock stability, the maximum being reached when $P_{3,0}=1$ which corresponds to a pure quantum state preparation. The same discussion applies to cold atom interferometers since their principle of operation is very similar to atomic clocks. The only difference being that the atom's motion is entangled with their internal quantum state in such a way that one can probe their motion by measuring the above mentioned clock transition probability. In this case, quantum state preparation will result in increased signal-to-noise ratio and thus increased sensor sensitivity. Other examples where quantum state preparation plays a crucial role include masers, lasers, and all quantum information science experiments.

Many different approaches have been used to prepare atomic samples in a given quantum state. The first thermal beam cesium clocks used Stern-Gerlach magnets to select atoms in one of the hyperfine ground states \cite{SternGerlachPaper}. Later, this selection process was advantageously replaced by laser optical pumping to transfer all the atoms from one hyperfine ground state to the other \cite{KastlerPaper,PicquePaper,ArditiPaper,OhshimaPaper,CerezPaper}. Then in the 1980s, two-laser optical pumping was proposed to produce both hyperfine and Zeeman pumping toward one of the clock state Zeeman sub-levels \cite{deClercqPaper,AvilaPaper,TremblayPaper}. However no improvement of the signal-to-noise ratio was observed because of the increased noise introduced by the pumping process \cite{LucasLeclinPaper,DimarcqPaper}. With the advent of cold atoms and atomic fountain clocks, two-laser optical pumping was abandoned because the number of pumping cycles is high ($>10$ on average with cesium) and thus produces an unwanted heating of the atomic cloud. Importance was put on purity of state preparation, and selection methods involving selective excitation followed by optical blowing of unwanted atoms were introduced \cite{FountainStatePreparationPaper}. Nowadays, the same methods are being used in cold atom interferometers \cite{InterferometerStatePreparationPaperChu,InterferometerStatePreparationPaperRasel,InterferometerStatePreparationPaperLandragin}. Many other original state preparation methods have been presented in the literature, for example optical pumping followed by rf transfer in a vapour cell \cite{BhaskarPaper,MicalizioPaper}, optical pumping via incoherent Raman transitions in cavity QED \cite{BoozerPaper}, preparation of pure superposition states with push-pull pumping \cite{JauPaper} or via elliptically polarized bichromatic fields \cite{ZibrovPaper}. 

In our experiment we want to prepare the quantum state of a continuous beam of cold atoms with a transverse temperature between 3 and 4~$\mu$K. Therefore we cannot afford the reheating due to spontaneous emission and, as a consequence, methods which combine state preparation and laser cooling are of particular interest for us. Moreover, we need a method which can be adapted to the continuous beam case. The first striking example of such combined internal and external state preparation was the observation of velocity selective coherent population trapping \cite{Aspect88} in 1988 with metastable helium atoms. Other schemes involving alkali-metal atoms include Zeeman-shift-degenerate Raman sideband cooling which prepares the atoms in the stretched state ($m=F$) \cite{HamannPaper,VuleticPaper,TreutleinPaper}. In Ref.~\cite{DiDomenicoSidebandPaper}, we showed that it is possible to adapt this scheme to a continuous beam of cold atoms. However, the resulting quantum state $\left|F=3,m=3\right\rangle$ is not useful for an atomic clock and thus we would need to replace in this sideband cooling scheme the Zeeman shift by a Stark shift to accumulate all the atoms in $\left|F=3,m=0\right\rangle$ as proposed in Ref.~\cite{TaichenachevPaper}. Even though this scheme seems very promising for cold atom clock applications, at present we are not aware of any experimental realization and we suppose that this may be due to the technical challenge of tailoring the required AC Stark shifts. As a first step toward this Stark-shift-degenerate Raman sideband cooling scheme, in this work we start by realizing optical pumping toward $m=0$ in parallel with Sisyphus cooling, similarly to the work described in Ref.\cite{ChoiPaper} which was realized with Doppler cooling in a 3D magneto-optical trap.

This article is organized as follows. In section~\ref{section2} we will start by presenting the principle of our optical pumping scheme and explain how one can combine both quantum state preparation and laser cooling in the same interaction zone. In section \ref{section3}, we will describe the experimental setup that we use to produce the continuous beam of cold atoms, to prepare the atoms in one of the clock states, and then to characterize both state preparation and laser cooling efficiency. In section \ref{section4} we will present our experimental results and discuss their physical interpretation in section \ref{section5}, and finally we will conclude in section \ref{section6}.

%%%%%%%%%%%%%%%%%%%%%%%%%%%%%%%%%%%%%%%%%%%%%%%%%%%%%%%%%%%%%%%%%%%%%%%%%%%%%
\section{Quantum state preparation principle}
\label{section2}
%%%%%%%%%%%%%%%%%%%%%%%%%%%%%%%%%%%%%%%%%%%%%%%%%%%%%%%%%%%%%%%%%%%%%%%%%%%%%

% Expliquer l'id\'ee principale: jouer avec la fr\'equence et la polarisation des lasers 
% pour produire du refroidissement qui compense le r\'echauffement. 
% L'utilisation d'un gradient de polarisation permet d'\'eviter les \'etats coh\'erents Zeeman. 
% La largeur de raie des lasers qui ne sont pas en phase ainsi que les longs temps d'interaction
% permettent conjointement d'\'eviter le pi\'egeage coh\'erent hyperfin.

Our objective is to prepare a continuous beam of cold cesium atoms in the ground state sub-level $\left|F=3,m=0\right\rangle$ by making use of two-laser optical pumping. A first laser is used to excite the $F=4$ ground state in order to pump the atoms into $F=3$ while a second $\pi$-polarized laser excites the $F=3\rightarrow F'=3$ transition. As a consequence, $\left|F=3,m=0\right\rangle$ is the only ground state sub-level which is not excited by laser light and therefore all atoms will accumulate into that state~\cite{deClercqPaper,AvilaPaper,TremblayPaper,LucasLeclinPaper,DimarcqPaper}. Therefore, the effect of the second laser is to produce Zeeman pumping toward $m=0$. Since we are working with a beam of cold atoms, we have to care about heating produced by the optical pumping cycles. Indeed, starting with a uniform population distribution among the $F=3,4$ ground states sub-levels, an average of more than ten optical pumping cycles are necessary to bring all the atoms into $\left|F=3,m=0\right\rangle$. Every pumping cycle transfers a random recoil to the atom and thereby increases the temperature of the atomic sample. In our case, this increase of temperature will result in an increased divergence of the atomic beam.

\begin{figure}[tbp]
\includegraphics[width=\grscale\textwidth]{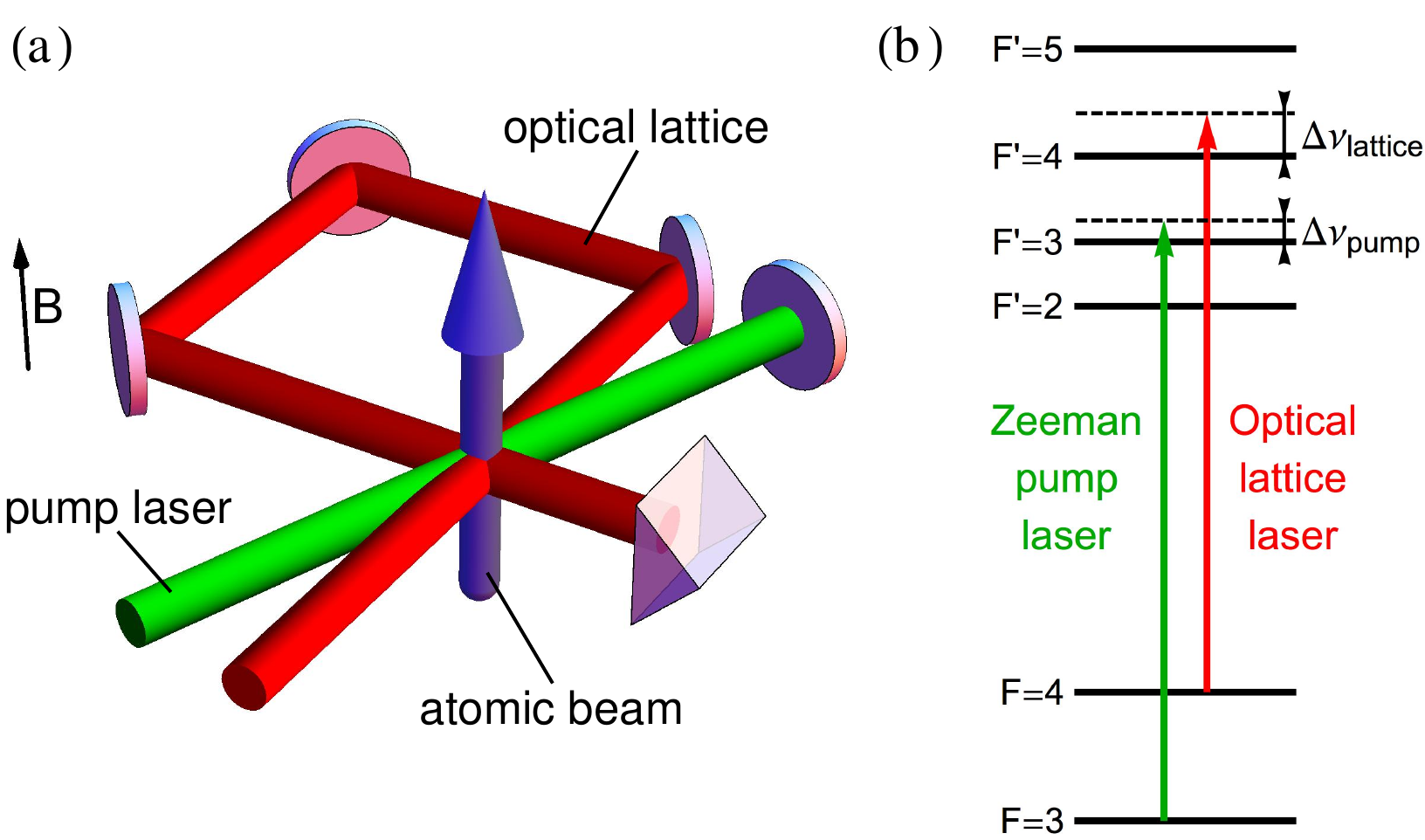}
\caption{(a) Configuration of laser beams used for simultaneous quantum state preparation and laser cooling. The atomic beam is vertical and all laser beams are in the same horizontal plane.  The optical lattice beam is folded and retro-reflected on a prism to create a phase-stable 2D optical lattice (see Ref.~\cite{LatticeStabilityPaper}). It has an incoming polarization which is linear at 45$^\circ$ with respect to the vertical direction, and subsequent multiple reflexions on metallic mirrors introduce some ellipticity. The Zeeman pump beam is linearly polarized in the vertical direction and retro-reflected with a mirror. It makes a small angle of approximately 5$^\circ$ with the lattice incoming direction. In principle this angle doesn't play any role, its value is simply restricted by the optical access to the vacuum system. All mirrors are metallic. A vertical magnetic field is used to stabilize the atomic polarization. (b) Frequencies of the laser beams with respect to the cesium D2 line transitions. In order to produce optical pumping toward $\left|F=3,m=0\right\rangle$ the optical lattice beam should excite either the $F=4\rightarrow F'=4$ or the $F=4\rightarrow F'=3$ transition, and the Zeeman pump beam should excite the $F=3\rightarrow F'=3$ transition.  As explained in the text, both lasers are detuned in order to produce laser cooling simultaneously with state preparation.}
\label{fig:Fig1}
\end{figure}

The main idea, which distinguishes this proposal from previous work, is to adjust the frequency, polarization, and geometry of the laser beams in such way as to produce Sisyphus cooling in order to compensate heating produced by the optical pumping cycles. The configuration of laser beams is shown in Fig.~\ref{fig:Fig1}(a) and their frequencies relative to the cesium D2 lines are defined in Fig.~\ref{fig:Fig1}(b). The atomic beam is vertical and all laser beams are in the same horizontal plane. 

The first laser beam, which produces hyperfine pumping from $F=4$ to $F=3$, is folded and retro-reflected on a prism to create a phase-stable 2D optical lattice (see Ref.~\cite{LatticeStabilityPaper}). Its incoming polarization is chosen linear at $45^\circ$ with respect to the vertical direction in such way as to produce a strong polarization gradient in the intersection region. With this polarization gradient and a carefully chosen laser frequency, one can observe Sisyphus cooling in the transverse directions. Another benefit of a polarization gradient is to avoid Zeeman coherent trapping states thanks to the motion of the atoms in the different polarization sites. In order to produce optical pumping toward $F=3$, the optical lattice beam should excite either the $F=4\rightarrow F'=4$ or the $F=4\rightarrow F'=3$ transitions, and by detuning its frequency on the blue side of those transitions, one can simultaneously produce Sisyphus cooling.

The second laser beam, which produces Zeeman pumping toward $m=0$, is linearly polarized in the vertical direction and retro-reflected with a mirror. Thanks to its polarization and according to selection rules, this laser will excite all sub-levels of $F=3$ except $m=0$. As a consequence, the atoms will tend to accumulate in $\left|F=3,m=0\right\rangle$. During this process, each time an atom is pumped back into $F=4$, it couples again to the optical lattice and thus restarts to be laser cooled. Therefore the global picture is as follows: either the atoms are in $F=3$ and they are pumped toward $m=0$, or they are in $F=4$ and they are laser cooled.

As a final remark, let's note that simultaneous excitation with two lasers detuned by the hyperfine frequency may, in principle, produce hyperfine coherent dark states. This would reduce the efficiency of state preparation and cause unwanted frequency shifts in the subsequent Ramsey resonance. However, in our experiment the two lasers are not phase locked and their linewidths are approximately $500$~kHz. These linewidths in conjunction with the long interaction time (a few ms) does not allow for the creation of coherent dark states.

%%%%%%%%%%%%%%%%%%%%%%%%%%%%%%%%%%%%%%%%%%%%%%%%%%%%%%%%%%%%%%%%%%%%%%%%%%%%%
\section{Experimental setup}
\label{section3}
%%%%%%%%%%%%%%%%%%%%%%%%%%%%%%%%%%%%%%%%%%%%%%%%%%%%%%%%%%%%%%%%%%%%%%%%%%%%%

%
\begin{figure}[tbp]
\includegraphics[width=\grscale\textwidth]{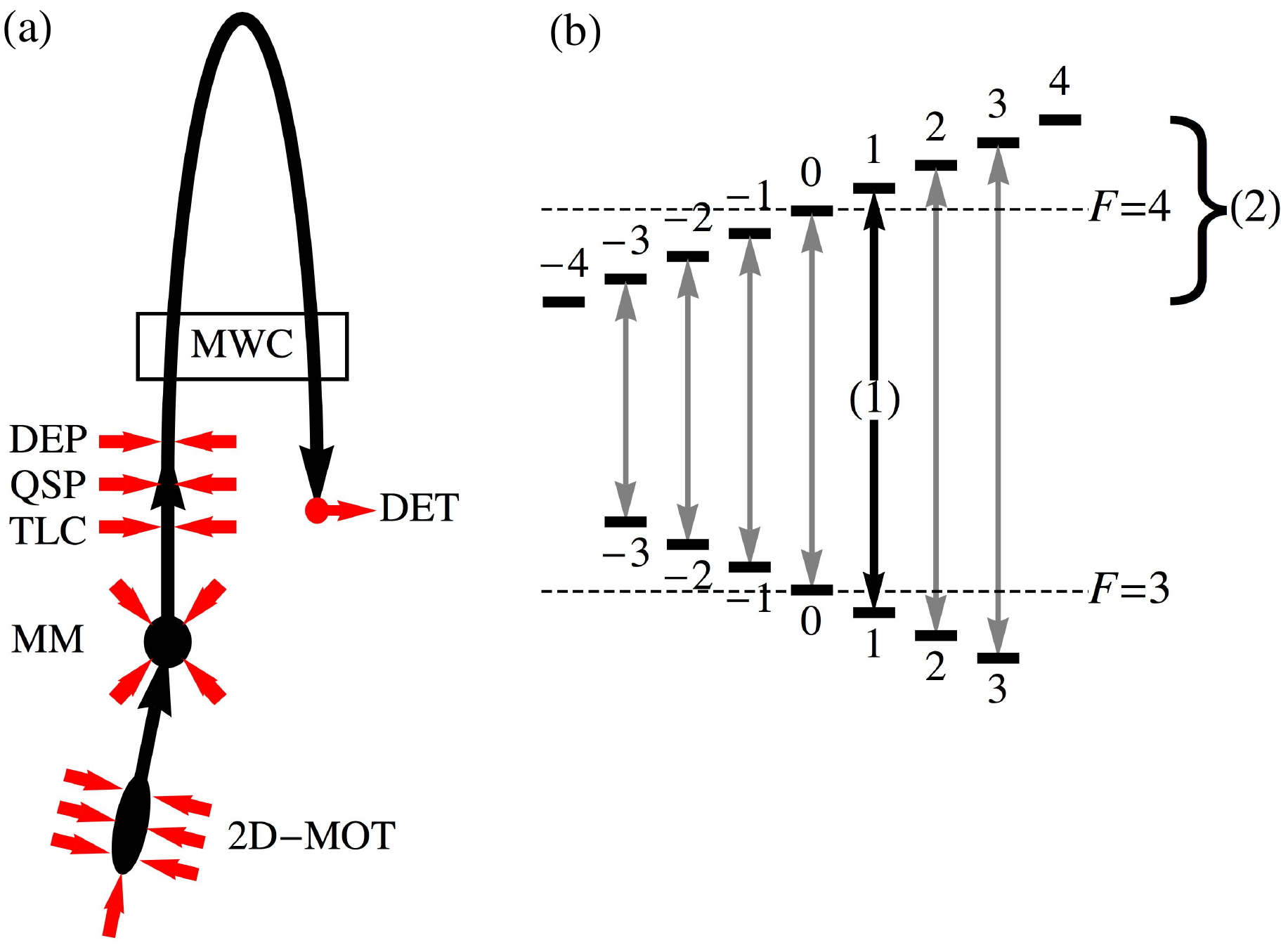}
\caption{(a) Scheme of the experimental setup. One can see the two-dimensional magneto-optical trap (2D-MOT) where the atomic beam is produced, the 3D moving molasses (MM) which cools and launches the atoms at a speed of $4$~m/s, the transverse laser cooling (TLC) which collimates the atomic beam, the quantum state preparation stage (QSP), the depumper (DEP), the microwave cavity (MWC), and finally the fluorescence detection (DET). (b) Scheme of Zeeman sub-level showing how the population distribution of the $F=3$ ground state is measured in two steps: (1) selective microwave excitation followed by (2) detection of atoms in $F=4$. See Section~\ref{section3} for details.}
\label{fig:Fig2}
\end{figure}

The scheme of the experiment is presented in Fig.~\ref{fig:Fig2}(a). It shows the main elements necessary to produce the continuous beam of cold cesium atoms, followed by the quantum state preparation region, the fountain parabola which passes two times through the microwave cavity, and finally the probe laser beam used for fluorescence detection.

The source of the atomic beam is a two-dimensional magneto-optical trap (2D-MOT) followed by a 3D moving molasses (MM) previously described in Refs~\cite{CastagnaPaper,PBe99} respectively. In the MM beams, the atoms are continuously captured, cooled and launched upward by the moving molasses technique. We thus obtain a continuous atomic beam with a total flux of $10^9$~at/s, a temperature between $50$ and $100$~$\mu$K and an
adjustable velocity, set at $4$~m/s in our experiment~\cite{ThomannPaper}. 

We further collimate the atomic beam in order to reduce the loss of atoms due to thermal expansion during the time of flight. To this end, we implement transverse laser cooling (TLC) in a two-dimensional optical lattice perpendicular to the atomic beam as described in Ref.~\cite{LatticeStabilityPaper}. By making use of Sisyphus cooling, we reduce the transverse temperature down to $4$~$\mu$K with an efficiency close to 100\%~\cite{DiDomenicoPhd}. At this level, the atomic beam has a diameter of approximately $10$~mm, and subsequently the effective fountain beam diameter is limited by the microwave cavity openings to $9$~mm.

Quantum state preparation (QSP) takes place after the collimation stage, $2.5$~cm above the TLC optical lattice. As described in Section~\ref{section2}, two lasers are necessary for the realization of our optical pumping scheme. The first laser, used for hyperfine pumping from $F=4$ to $F=3$, is implemented as a folded optical lattice as shown in Fig.~\ref{fig:Fig1}(a). Its power ($2.5$~mW) and frequency ($125$~MHz above the $F=4\rightarrow F'=4$ transition) are chosen to optimize the cooling, and therefore limit the heating produced by the optical pumping cycles. Its incoming polarization is chosen linear at $45^\circ$ with respect to the vertical direction in order to produce a strong polarization gradient, with the double benefit to enable Sisyphus cooling and avoid the formation of Zeeman coherent states at the same time. The second laser, used for Zeeman pumping to the $\left|F=3,m=0\right\rangle$ ground state, is superposed upon the optical lattice as described in Fig.~\ref{fig:Fig1}(a). It is retro-reflected to minimize pushing of the atomic beam. In the basic configuration, its frequency is tuned on resonance with the $F=3\rightarrow F'=3$ transition and its power ($3$~$\mu$W) has been adjusted to optimize the flux of atoms in $\left|F=3,m=0\right\rangle$. Moreover, we observed that it is possible to improve quantum state preparation by detuning this laser a few MHz above the $F=3\rightarrow F'=3$ transition and  adjusting its power accordingly. This will be discussed in detail in Sections~\ref{section4} and~\ref{section5}. Both the optical lattice and Zeeman pump laser beams are gaussian with a waist of $5.7$~mm and truncated at a diameter of $11$~mm. With an atomic beam velocity of $3.6$~m/s through the state preparation region we obtain a transit time of $3$~ms. According to our numerical simulations, this transit time is sufficient to produce an almost complete (98\%) inversion of the clock transition.

A crucial point of this experiment concerns the state and purity of the Zeeman pump laser polarization. Indeed, to be effective, $\pi$-pumping toward $m=0$ requires a linear polarization aligned with the quantization axis which is determined by the magnetic field direction. Therefore, not only the direction of the polarization vector has to be finely tuned to that of the magnetic field, but also any ellipticity should be avoided. Since the laser light is transported to the vacuum chamber with polarization maintaining fibers, special care was taken to optimize the extinction ratio, which was measured to be $40$~dB after the first passage through the vacuum chamber i.e. before the retro-reflection mirror. Any component of the magnetic field perpendicular to the laser polarization destroys the atomic alignment. Therefore a small component of the magnetic field parallel to the laser polarization is necessary to stabilize the created atomic alignment against the depolarization effect of the unavoidable residual magnetic field inhomogeneities in the transverse directions. In our experiment, we use three pairs of coils mounted in Helmoltz-like configuration to control the magnetic field in this region. The external fields are thus compensated and one can finely tune the value and direction of the resulting magnetic field ($\approx 1$~$\mu$T) to align it with the pump laser polarization.

After state preparation, a small fraction of the atoms (of the order of $10$\%) remain in the $F=4$ ground state, probably due to experimental imperfections like the difficulty to perfectly superpose the optical lattice and Zeeman pump lasers, stray light scattered and reflected by the windows of the vacuum system, and fluorescence light from the laser cooling regions. Therefore we use a depumper laser (DEP) to completely depopulate the $F=4$ ground state before performing Ramsey spectroscopy. This laser is tuned on the $F=4\rightarrow F'=4$ transition with a power of $0.8$~mW. It is sent perpendicularly to the atomic beam, $2$~cm above the state preparation stage, and retro-reflected. Note that all these laser beams (TCL, QSP and DEP) are tilted by $1.6^\circ$ with respect to the horizontal, in such way that the atoms describe an open parabola passing through the microwave cavity before reaching the detection region.

In order to characterize the state preparation efficiency, we use Ramsey spectroscopy to measure the population distribution on the Zeeman sub-levels of the $F=3$ ground state. The measurement principle, illustrated in Fig.~\ref{fig:Fig2}(b), proceeds in two steps: firstly the population of $\left|F=3,m\right\rangle$ is transferred into $\left|F=4,m\right\rangle$ by selective microwave excitation, and then the $F=4$ population is measured by fluorescence detection. More precisely, the microwave excitation exchanges the populations of $\left|F=3,m\right\rangle$ and $\left|F=4,m\right\rangle$, and thus this method gives a measurement of $P_{3,m}-P_{4,m}$ where $P_{F,m}$ denotes the population of state $\left|F,m\right\rangle$. However, thanks to the depumper laser, the total population of $F=4$ is smaller than $3~$\%. Therefore, with the resonable hypothesis that at the end of the two-laser pumping process the population distribution of both ground states are similar ($P_{4,m}\approx 0.03 P_{3,m}$), this method gives an estimate of $P_{3,m}$ with a relative error smaller than $3$\%. In practice, the two Rabi interactions are spatially separated in our continuous fountain, and the atoms pass through the microwave cavity in the upward direction, having a first $\pi/2$-pulse, freely evolve for $\approx 0.5$s and turn back into the cavity where the second $\pi/2$-pulse is applied. The transit time in each Ramsey interrogation zone is approximately $10$~ms. A vertical magnetic field, the so called C-field, is used to lift the degeneracy of the seven $\left|F=3,m\right\rangle \rightarrow \left|4,m\right\rangle$ transitions as shown in Figure \ref{fig:Fig2}(b). The experimental value of $77$~nT shifts the seven resonances by $m\times 540$~Hz around the $\left|F=3,0\right\rangle \rightarrow \left|F=4,0\right\rangle$ clock transition. By scanning the microwave frequency around each transitions, one can then probe their population by fluorescence detection of the $F=4$ atomic flux. To this end, a retro-reflected probe laser beam ($14$~mm diameter and $1$~mW of power) tuned $2$ - $5$~MHz below the $F=4\rightarrow F'=5$ transition is sent through the atomic beam, and the fluorescence light is collected and measured with a low noise photo-detector (DET).

Finally, by superposing a repumper laser (30~$\mu$W tuned to the $F=3\rightarrow F'=4$ transition) on the probe laser, we can detect all the atoms in both ground states and thus obtain a measure of the total fountain flux. In our experiment, we use the total flux as an indirect measurement of the cooling efficiency of quantum state preparation. Indeed, the fountain flux is reduced by the losses due to the thermal expansion of the atomic beam during the time of flight. Since the atomic beam section expands proportionally to the square of the transverse velocity, the total fountain flux is inversely proportional to the transverse temperature. This  will be discussed in more detail in Section~\ref{section5}.

%%%%%%%%%%%%%%%%%%%%%%%%%%%%%%%%%%%%%%%%%%%%%%%%%%%%%%%%%%%%%%%%%%%%%%%%%%%%%
\section{Experimental results}
\label{section4}
%%%%%%%%%%%%%%%%%%%%%%%%%%%%%%%%%%%%%%%%%%%%%%%%%%%%%%%%%%%%%%%%%%%%%%%%%%%%%

%%%%%%%%%%%%%%%%%%%%%%%%%%%%%%%%%%%%%%%%%%%%
\subsection{Efficiency of state preparation}
%%%%%%%%%%%%%%%%%%%%%%%%%%%%%%%%%%%%%%%%%%%%

%
\begin{figure}[tbp]
\includegraphics[width=\grscale\textwidth]{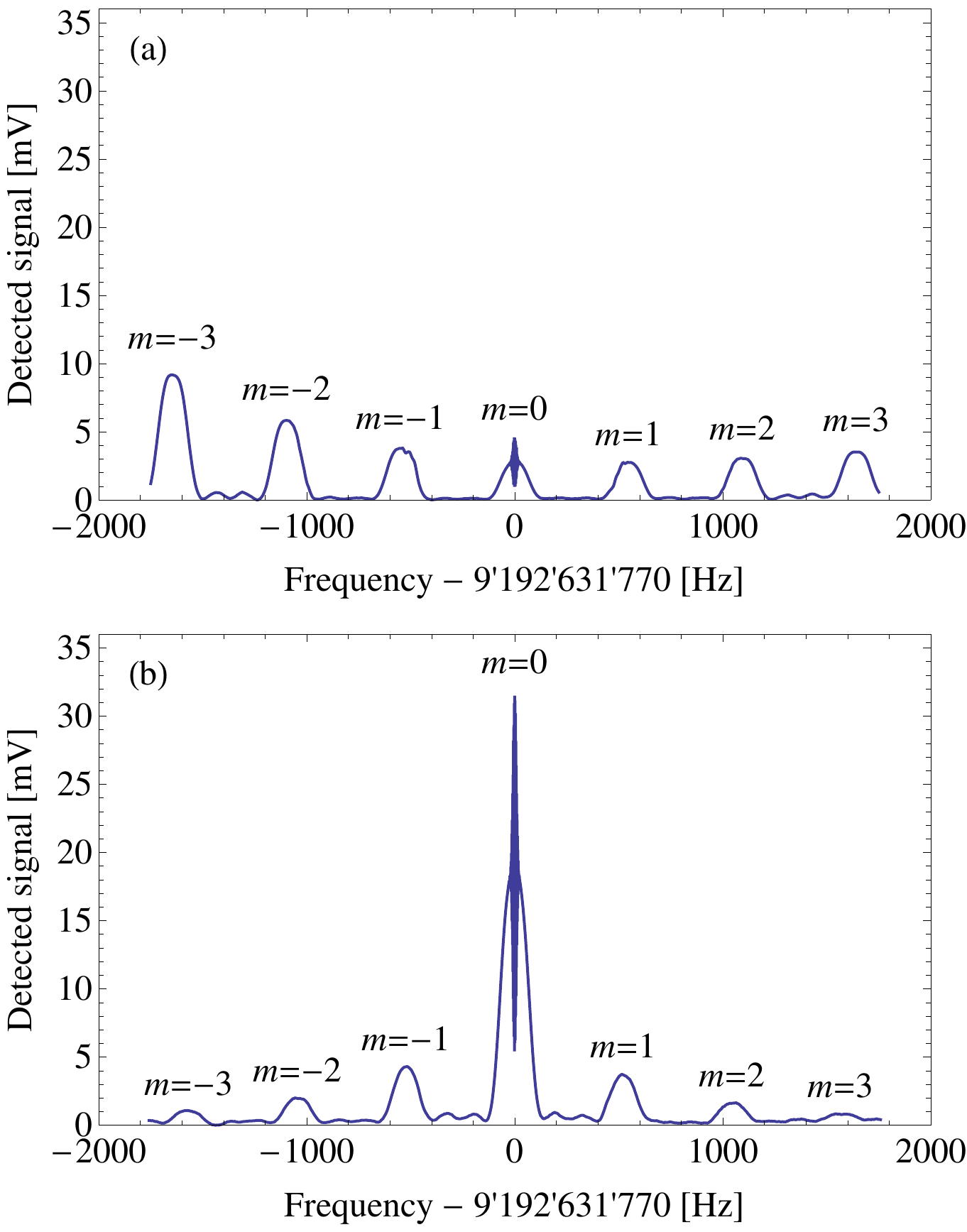}
\caption{Microwave spectra measured (a) without state preparation and (b) with state preparation. These spectra represent the number of atoms detected in $F=4$ as a function of the microwave frequency. When the microwave is resonant with one Zeeman component ($\left|3,m\right\rangle\rightarrow \left|4,m\right\rangle$) atoms are transferred in $F=4$ and detected. The microwave power is adjusted for each Zeeman component in order to produce two $\pi/2$-pulses. Note that we used identical scales for both graphs to facilitate the comparison. }
\label{fig:Fig3}
\end{figure}

The microwave spectra measured with and without state preparation are presented in Fig.~\ref{fig:Fig3}. They were obtained by scanning the microwave frequency and measuring the number of atoms detected in $F=4$. As explained in section~\ref{section3}, before the microwave interrogation a depumper is used to depopulate the $F=4$ ground state, and the microwave power is optimized for each Zeeman component in order to produce two $\pi/2$-pulses. As a consequence, the Rabi envelope gives a good approximation (error $<3$\%) of the $F=3$ population distribution. 

The microwave spectrum obtained without state preparation is shown in Fig.~\ref{fig:Fig3}(a). It is composed of seven Rabi resonances corresponding to the transitions $\left|3,m\right\rangle\rightarrow \left|4,m\right\rangle$ for $m=-3,...,+3$. 
Ramsey fringes are clearly visible on the central resonance ($m=0$) but not on the other Zeeman components. This is partly due to magnetic field inhomogeneities in the microwave interrogation region but mostly to the frequency sampling resolution which was adjusted to observe the Rabi resonances. Indeed, for practical reasons the frequency is sampled every $4$~Hz except in a band of $\pm 25$~Hz around $9192631770$~Hz where the resolution is increased to $0.2$~Hz in order to observe the central Ramsey fringes. Remark that $4$~Hz resolution is far enough to measure the Rabi resonances which have a width of approximately $150$~Hz. As a result, one observes in Fig.~\ref{fig:Fig3}(a) that the population distribution among the Zeeman sub-levels of $F=3$ is not uniform, it is asymmetric, and only 8.7\% of the atoms are in $m=0$ (see also Table~\ref{tab1}). 

Both the non-uniformity and the asymmetry of the population distribution may be explained by the polarization of the folded optical lattice used for transverse cooling which contains some ellipticity induced the multiple reflexions on metallic mirrors. For our application, namely a primary frequency standard, both the asymmetry and the small number of atoms in $m=0$ are problematic. Indeed, the asymmetry may produce Rabi and Ramsey pulling \cite{VanierAudoinBook}, and the small atom number will reduce the signal-to-noise ratio and thus degrade the clock stability. 

These two problems can be solved simultaneously by introducing state preparation. Indeed, the microwave spectrum measured with state preparation is shown in Fig.~\ref{fig:Fig3}(b). On this graph, one can see that the population distribution is quite symmetric and that 56.6\% of the atoms accumulated in $m=0$. More precisely and for comparison, the populations obtained from both microwave spectra are summarized in Table~\ref{tab1}. In order to quantify the gain in symmetry, we calculated the orientation from both population distributions. 

Finally, we can conclude that state preparation decreased the orientation (asymmetry) by a factor twelve, and it increased the population of $m=0$ by a factor six. In section~\ref{section5}, we will discuss with more detail the factors limiting the purity of state preparation in our experiment. 

\begin{table}[tbp]
\begin{ruledtabular}
\begin{tabular}{lcc}
          & Without   & With  \\
        & state preparation & state preparation \\
  \hline
  $P_{3,-3}$ & $29.4$~\% & $3.5$~\% \\
  $P_{3,-2}$ & $18.9$~\% & $6.6$~\% \\
  $P_{3,-1}$ & $12.5$~\% & $13.5$~\% \\
  $P_{3,0}$ & $8.7$~\% & $56.6$~\% \\
  $P_{3,1}$ & $9.1$~\% & $11.6$~\% \\
  $P_{3,2}$ & $10.1$~\% & $5.2$~\% \\
  $P_{3,3}$ & $11.3$~\% & $3.0$~\% \\
  \hline
  Orientation &  $-25.1$~\% & $-2.1$~\% \\
%  Alignment   &   $21$~\% & $-67$~\% \\
\end{tabular}
\end{ruledtabular}
  \caption{Populations of the $F=3$ ground state Zeeman sub-levels measured with and without state preparation (data obtained from Fig.~\ref{fig:Fig3}). Orientation is given by ${\cal O}\propto\Sigma m P_{3,m}$ and it is normalized relative to its maximum value. It is a measure of the population distribution dipole moment and thus of its asymmetry. 
% Alignment is given by ${\cal A}\propto \sum (3m^2-F(F+1)) P_{3,m}$. It is a measure of the population distribution
% quadrupole moment. Both orientation and alignment are normalized relative to their minimum and maximum values. 
Perfect state preparation should give $P_{3,0}=100$~\% and $\mathrm{orientation}=0$~\%.}
  \label{tab1}
\end{table}

%%%%%%%%%%%%%%%%%%%%%%%%%%%%%%%%%%%%%%%%%%%%
\subsection{Evidence of laser cooling during state preparation}
%%%%%%%%%%%%%%%%%%%%%%%%%%%%%%%%%%%%%%%%%%%%

\begin{figure}[tbp]
\includegraphics[width=\grscale\textwidth]{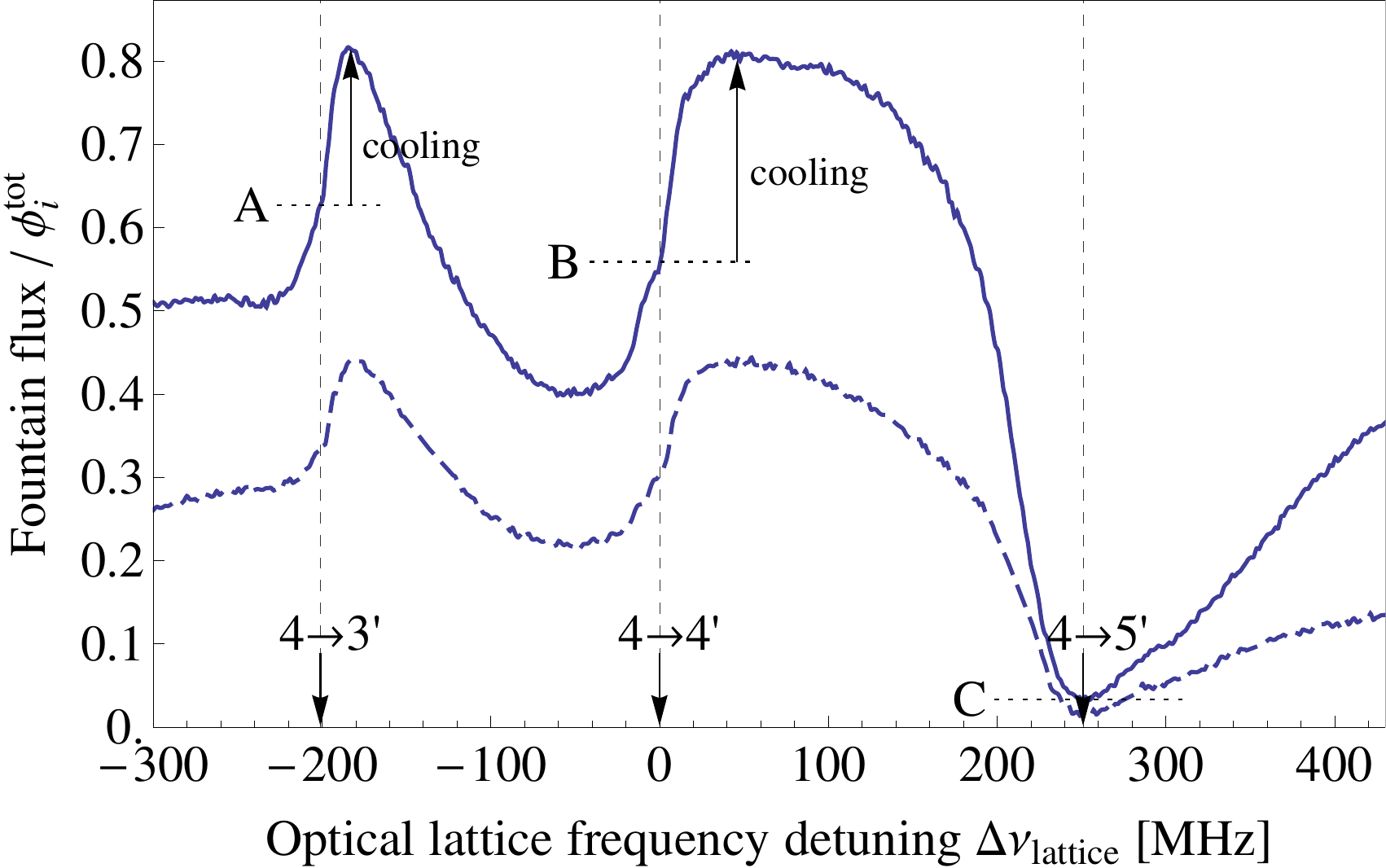}
\caption{Measurement of the total fountain flux in $F=3,4$ (solid line) and the flux in $\left|F=3,m=0\right\rangle$ (dashed line) as a function of the optical lattice laser frequency. $\Delta\nu_\mathrm{lattice}$ is the lattice frequency detuning from the $F=4\rightarrow F'=4$ transition of cesium D2 line. The Zeeman pump laser is tuned on the $F=3\rightarrow F'=3$ transition. The vertical axis is normalized to the total flux obtained without state preparation $\phi^\mathrm{tot}_{i}$. The dotted horizontal lines A, B and C indicate the total flux levels when the optical lattice laser is on resonance with the three transitions respectively. Cooling is observed on the blue side of those transitions, see text for details.}
\label{fig:Fig4}
\end{figure}

In order to observe laser cooling we scanned the optical lattice laser frequency and measured the total flux at the end of the fountain i.e. in the detection zone. As explained in section~\ref{section3}, a decrease of the atomic beam transverse temperature will lead to a decrease of the losses during the parabolic flight and thereby to a higher flux measured in the detection zone. More precisely, the total flux is inversely proportional to the transverse temperature and therefore it can be used as an indirect temperature measurement. During this measurement we recorded both the total flux ($F=3,4$) and the flux in $\left|F=3,m=0\right\rangle$ and the results are shown in Fig.~\ref{fig:Fig4}. One can see that the total flux does not change symmetrically around the $F=4\rightarrow F'=3$ and $F=4\rightarrow F'=4$ transitions. It increases on the blue side, which is a signature of Sisyphus cooling, and decreases on the red side, which is a signature of Sisyphus heating. This point will be discussed in more detail in section~\ref{section5}. Finally, we note that approximately 55\% of the atoms are in $m=0$ over a large range of frequencies. This indicates that state preparation is not limited by the optical lattice hyperfine pumping rate and that it takes place independently of the laser cooling processes.

%

%%%%%%%%%%%%%%%%%%%%%%%%%%%%%%%%%%%%%%%%%%%%
\subsection{Role of the pump laser frequency}
%%%%%%%%%%%%%%%%%%%%%%%%%%%%%%%%%%%%%%%%%%%%

\begin{figure}[tbp]
\includegraphics[width=\grscale\textwidth]{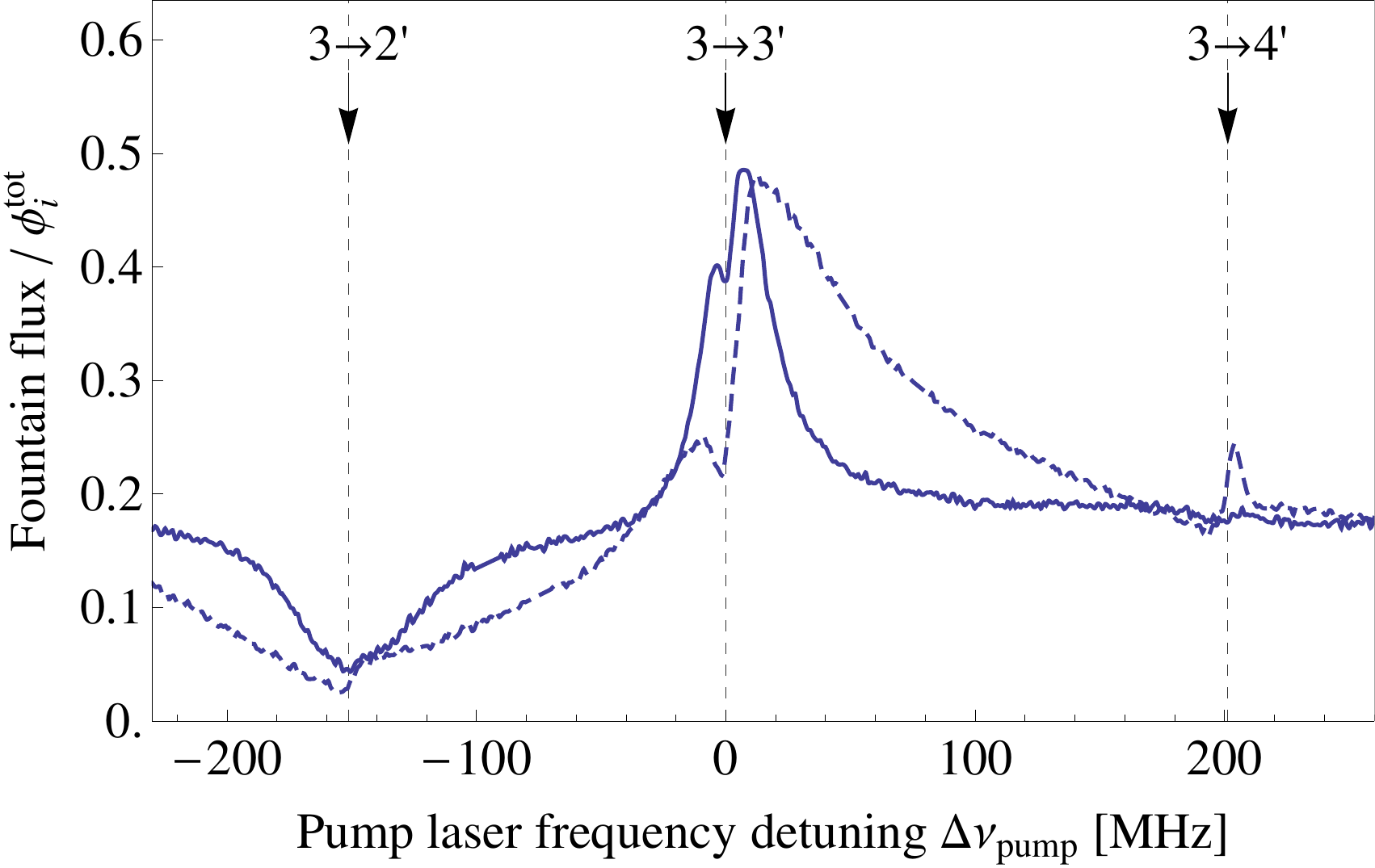}
\caption{Measurement of the $\left|F=3,m=0\right\rangle$ fountain flux as a function of the Zeeman pump laser frequency. $\Delta\nu_\mathrm{pump}$ is the pump laser frequency detuning from the $F=3\rightarrow F'=3$ transition of cesium D2 line. The pump laser power values are 3~$\mu$W (solid line) and 30~$\mu$W (dashed line). The vertical axis is normalized to the total flux obtained without state preparation $\phi^\mathrm{tot}_{i}$. The horizontal axis is calibrated using a saturated absorption signal obtained from a small fraction of the pump laser.}
\label{fig:Fig5}
\end{figure}

In principle, the role of the pump laser should be limited to optical pumping and thus we expected a frequency dependence symmetric around the atomic transitions, in contrast with the optical lattice laser. However, the interplay between two lasers may bring surprises and therefore we repeated the measurement of previous section but this time by scanning the pump laser frequency and measuring the fountain flux in $\left|F=3,m=0\right\rangle$. The results are presented in Fig.~\ref{fig:Fig5} for two different values of the pump power. 

Firstly, we observe that the $m=0$ fountain flux is increased on the $3\rightarrow 3'$ transition because $\pi$-optical pumping populates $m=0$, and decreased on the $3\rightarrow 2'$ transition because $\pi$-optical pumping populates $m=\pm 3$. Secondly, one can see that the $m=0$ flux can be notably increased by detuning the pump laser a few MHz to the blue side of the $3\rightarrow 3'$ transition. It is important to emphasize that the asymmetry observed around the $3\rightarrow 3'$ transition cannot be explained by optical pumping mechanisms only. The light shift produced by the pump laser should be accounted for in order to explain such an asymmetry. Our interpretation is that we observe the first signs of Stark-shift-degenerate Raman sideband cooling as proposed in Ref.~\cite{TaichenachevPaper}. This point will be discussed further in section~\ref{section5}. 

We repeated this measurement for a higher power of the pump laser and the maximum of the $m=0$ flux is shifted toward higher frequencies. For a pump power of $250$~$\mu$W this maximum approaches $25$~MHz above the $3\rightarrow 3'$ transition, which is interesting for a practical implementation of this scheme (laser locking) since it coincides with the  saturated absorption cross-over between $3\rightarrow 2'$ and $3\rightarrow 4'$ transitions. By locking the pump laser to this point, we observed that laser cooling compensates completely the heating produced by state preparation.

Finally, we observe a small increase of the $m=0$ flux on the $3\rightarrow 4'$ transition. According to our simulations, this is due to a concentration of the atoms in the low $m$ values provoked by this optical pumping configuration. 

% First of all, we observe some heating (decrease of the total flux) on each atomic transition which is due to optical
% pumping cycles, only partially compensated by Sisyphus cooling, partly because the optical lattice laser was locked midway
% between the $4\rightarrow 4'$ and $4\rightarrow 5'$ transitions for practical reasons. 

%
%

%%%%%%%%%%%%%%%%%%%%%%%%%%%%%%%%%%%%%%%%%%%%
\subsection{Atomic beam noise measurement}
%%%%%%%%%%%%%%%%%%%%%%%%%%%%%%%%%%%%%%%%%%%%

As mentioned in the introduction, two-laser optical pumping was implemented for state preparation in a thermal cesium beam resonator but a degradation of the signal-to-noise ratio was observed due to the presence of excess noise on the fluorescence signal \cite{DimarcqPaper}. Later this additional noise was analyzed and attributed to the presence of residual unpumped atoms combined with frequency fluctuations of the pumping laser \cite{LucasLeclinPaper}. Given that our experimental conditions are very different (lower atomic flux, longer interaction time, low frequency noise lasers) our hope was to observe an improvement of the signal-to-noise ratio by introducing state preparation. 

\begin{figure}[tbp]
\includegraphics[width=\grscale\textwidth]{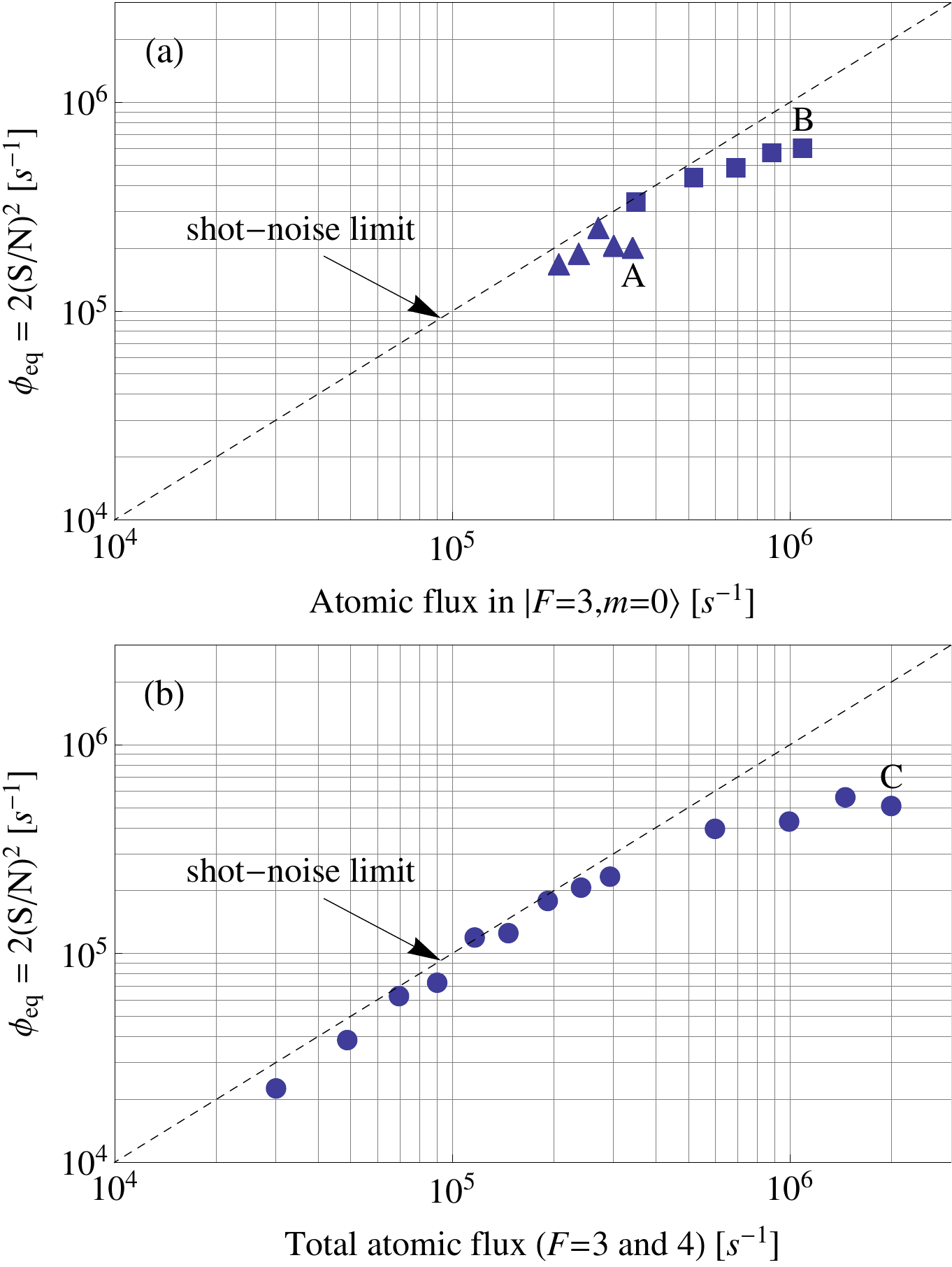}% two-column mode
\caption{Evolution of $\phi_{eq}=2(S/N)^2$ where $S$ is the detected signal and $N$ its noise spectral density, as a function of the atomic flux. (a) The triangles are measurements of the atoms in $\left|F=3,m=0\right\rangle$ obtained without state preparation. The squares are measurements of the atoms in $\left|F=3,m=0\right\rangle$ obtained with state preparation. (b) The circles are measurements of the total flux in $F=3$ and $4$. As explained in the text, the fountain is shot-noise limited at low flux and thus allows a calibration of the photodetector signal in atoms$/$sec (horizontal axis). On these graphs the shot-noise limit is a line with unit slope. Note that the three points marked by letters A, B, C were obtained with the maximum total fountain flux.}
\label{fig:Fig6}
\end{figure}

In order to demonstrate that, we measured the signal-to-noise ratio of the $m=0$ fountain flux with and without state preparation. More precisely, in our measurement the signal $S$ ($\mathrm{A}$) is obtained from the DC current of the fluorescence detection photodiode, and the noise $N$ ($\mathrm{A}/\sqrt{Hz}$) from the linear spectral density of the photodiode current at $1$~Hz (the fountain clock modulation frequency). In order to observe any departure from the atomic shot noise level, we repeated the measurement for different values of the total flux. Given that the fountain is shot-noise limited at low flux, this measurement has the advantage to allow for an absolute calibration of the detection efficiency \cite{GuenaPaper,DiDomenicoPhd}. The results are presented in Fig.~\ref{fig:Fig6} where we reported $\phi_{eq}=2(S/N)^2$ as a function of the atomic flux. Here $\phi_{eq}$ is the shot-noise limited equivalent flux i.e. which would give the same signal-to-noise ratio.

One can see in Fig.~\ref{fig:Fig6} that the signal-to-noise ratio is improved thanks to state preparation. Indeed, with state preparation (point B) we measure a threefold improvement of the maximum equivalent flux compared to the situation without state preparation (point A). The graph also shows a departure from the shot-noise limit for higher values of the flux. However this behavior is also observed on the total flux and thus cannot be attributed to  state preparation. Indeed, the maximum $S/N$ of the $m=0$ flux with state preparation (point B) is approximately equal to the maximum $S/N$ of the total flux (point C). Therefore, our interpretation is that this noise is imprinted on the atomic flux during the laser cooling stages before state preparation.

%

%%%%%%%%%%%%%%%%%%%%%%%%%%%%%%%%%%%%%%%%%%%%%%%%%%%%%%%%%%%%%%%%%%%%%%%%%%%%%
\section{Discussion of results}
\label{section5}
%%%%%%%%%%%%%%%%%%%%%%%%%%%%%%%%%%%%%%%%%%%%%%%%%%%%%%%%%%%%%%%%%%%%%%%%%%%%%

%%%%%%%%%%%%%%%%%%%%%%%%%%%%%%%%%%%%%%%%%%%%
\subsection{What limits state preparation purity}
%%%%%%%%%%%%%%%%%%%%%%%%%%%%%%%%%%%%%%%%%%%%

% Qu'est-ce qui limite la purete de la preparation d'etat. Discuter les conditions experimentales. 
% Inhomogeneite du champ magnetique. Les simulations confirment. Interpretation physique?

As reported in Table~\ref{tab1}, we were able to accumulate 56.6\% of the atoms in $\left|F=3,m=0\right\rangle$ with our state preparation scheme. This is a sixfold improvement over the situation without state preparation, however this scheme should allow us to reach 100\% in $m=0$, at least in principle. In order to understand what is the limiting factor in our experiment, we developed a numerical model of optical pumping based on the rate equations presented in Ref.~\cite{AvilaPaper} with the notable difference that we took into account off-resonance excitation of all transitions. 

We performed numerical simulations of state preparation with the same parameters as in the experiment. The optical lattice laser frequency is $125$~MHz above the $F=4\rightarrow F'=4$ transition and has a power of $2.5$~mW per beam (i.e. $2.5$~mW/cm$^2$ on average). The Zeeman pump laser frequency is on resonance with the $F=3\rightarrow F'=3$ transition and it has a power of $3$~$\mu$W (i.e. $3$~$\mu$W/cm$^2$ on average). The length of the state preparation zone, where all the laser beams are superposed to the atomic beam, is estimated to be of $11$~mm. With an atomic beam velocity of $3.6$~m/s, we obtain a transit time of $3$~ms. The laser light is transported to the vacuum system with polarization maintaining fibers. The extinction ratio of the pump laser beam was optimized with care and was measured to be $40$~dB. Therefore the intensity of the circular polarization component of the pump beam is smaller than $1/10000$ of the total intensity. The homogeneity of magnetic field was more difficult to control in this region of the experiment. In order to stabilize the atomic polarization, we create a vertical magnetic field of $1$~$\mu$T with rectangular coils in Helmoltz-like configuration. However, the presence of $\mu$-metal material close to this region (the magnetic shields of the interaction zone) deforms significantly the magnetic field lines and thus the field homogeneity is difficult to evaluate. 
% As a consequence, we measured that the horizontal component of the magnetic field varies 
% by $\pm 6$\% of the total field over the state preparation volume.
After inserting these experimental parameters in our optical pumping model, we found that $6$\% of transverse residual magnetic field can explain our experimental results. Indeed, the populations obtained from our simulations are in good agreement with the populations measured in Table~\ref{tab1} with state preparation. More precisely, the populations are 59.5\% in $m=0$, 11.7\% in $m=\pm 1$, 5.8\% in $m=\pm 2$, and 2.7\% in $m=\pm 3$. Therefore the magnetic field inhomogeneity is probably the main experimental factor limiting the state preparation purity in our experiment.

%%%%%%%%%%%%%%%%%%%%%%%%%%%%%%%%%%%%%%%%%%%%
\subsection{Laser cooling compensates the heat produced by state preparation}
%%%%%%%%%%%%%%%%%%%%%%%%%%%%%%%%%%%%%%%%%%%%

% Discuter le r\'echauffement provoqu\'e par la pr\'eparation d'etat avec un petit calcul. 
% Argumenter qu'on observe du refroidissement/r\'echauffement Sisyphe dans le bleu/rouge des transitions 43 et 44. 
% Comparer 43 et 44 qui ont tendance a accumuler les atomes au centre respectivement dans les bords.

State preparation with optical pumping heats the atomic beam due to the random recoils generated by spontaneous emission during optical pumping cycles. In order to understand the role of laser cooling, we calculated the number of spontaneously emitted photons generated by state preparation when both the Zeeman pump and the optical lattice lasers are on resonance. By numerical integration of rate equations, and with the same conditions as in the experiment, we obtain the numbers summarized in Table~\ref{tab2}. From the number of recoils, we calculated the elevation of transverse temperature of the atomic beam. The initial value of $T_{i}=4$~$\mu$K was measured in Ref.~\cite{DiDomenicoPhd} and the final value $T_{f}$ is calculated according to statistical kinetic theory:
\begin{equation}
\label{Eq1}
	\frac{1}{2}k_B T_{f} = \frac{1}{2}k_B T_{i} + \frac{1}{2}M v_r^2 N_\mathrm{photons}/3
\end{equation}
where $k_B$ is the Boltzmann constant, $v_r=3.5\cdot 10^{-3}$~m/s is the recoil velocity and $M$ the cesium atomic mass. From the elevation of the atomic beam transverse temperature we calculated the decrease of the detected fountain flux according to $\phi^\mathrm{tot}_{f}/\phi^\mathrm{tot}_{i}=T_{i}/T_{f}$ and the results are summarized in Table~\ref{tab2}. These calculated fluxes ratios are in good agreement with the levels indicated by the horizontal lines A, B and C in the measurement of Fig.~\ref{fig:Fig4}. They give us reference levels with respect to which laser cooling can be observed in the measurement of Fig.~\ref{fig:Fig4}.

Now let's discuss the optical lattice frequencies at which we can expect laser cooling to operate in our experiment. A condition for Sisyphus cooling to work is that optical pumping always populates the Zeeman sub-levels with minimum energy. According to transitions oscillator strengths, this condition is fulfilled on the blue side of transitions $F\rightarrow F'\le F$ and on the red side of transitions $F\rightarrow F'>F$. Indeed, Sisyphus cooling was demonstrated to work for a positive detuning when $F'\le F$ and for a negative detuning when $F'>F$ \cite{drewsen96,Ellmann01}. This is exactly what we observe in Fig.~\ref{fig:Fig4}: cooling operates on the blue side of transitions $4\rightarrow 3'$ and $4\rightarrow 4'$, but on the red side of transition $4\rightarrow 5'$. Let's remark that in the first case, i.e. $F'\le F$ and positive detuning, atoms are pumped toward states weakly coupled to laser light (gray optical lattice) and thus heating by photon scattering is reduced. While in the second case, i.e. $F'>F$ and negative detuning, atoms are pumped toward states maximally coupled to laser light (bright optical lattice) and thus photon scattering is increased. This explains that the minimum temperatures, corresponding to maximum total fluxes, are observed close to the $4\rightarrow 3'$ and $4\rightarrow 4'$ transitions.

To conclude this discussion, let's emphasize that the atoms entering the state preparation region are already cold ($T_{i}=4$~$\mu$K) but they are heated by the optical pumping cycles necessary for state preparation. In this situation, we observe that Sisyphus cooling removes most of the heat produced by state preparation, and therefore help us to take the best advantage of state preparation.

\begin{table}[tbp]
\begin{ruledtabular}
\begin{tabular}{lccc}
  Optical lattice frequency & $4\rightarrow 3'$ & $4\rightarrow 4'$ & $4\rightarrow 5'$ \\
  \hline
  $N_\mathrm{photons}$ &  $37$ & $44$ & $31000$  \\
  $T_{i}$ ($\mu$K) & $4$  & $4$  & $4$      \\
  $T_{f}$ ($\mu$K) & $6.4$  & $6.9$  & $2040$      \\
  $\phi^\mathrm{tot}_{f}/\phi^\mathrm{tot}_{i}$ & 0.62 & 0.58 & 0.002     \\
\end{tabular}
\end{ruledtabular}
  \caption{Estimation of the elevation of temperature produced by state preparation lasers when the optical lattice is exactly on resonance with cesium D2 line transitions. We calculate the number of photons $N_\mathrm{photons}$ emitted spontaneously during state preparation by numerical integration of the rate equations. $T_{i}$ is the atomic beam transverse temperature before state preparation measured in previous work~\cite{DiDomenicoPhd}. $T_{f}$ is the temperature after state preparation calculated from $T_{i}$ and $N_\mathrm{photons}$ using Eq.~\ref{Eq1}. $\phi^\mathrm{tot}_{f}/\phi^\mathrm{tot}_{i}$ is the ratio of the total fountain flux with and without state preparation lasers. This calculated ratio is in good agreement with the measured levels A,B and C shown in Fig.~\ref{fig:Fig4}}
  \label{tab2}
\end{table}

%%%%%%%%%%%%%%%%%%%%%%%%%%%%%%%%%%%%%%%%%%%%
\subsection{Effect of the Zeeman pump laser frequency on state preparation}
%%%%%%%%%%%%%%%%%%%%%%%%%%%%%%%%%%%%%%%%%%%%

%
\begin{figure}[tbp]
\includegraphics[width=\grscale\textwidth]{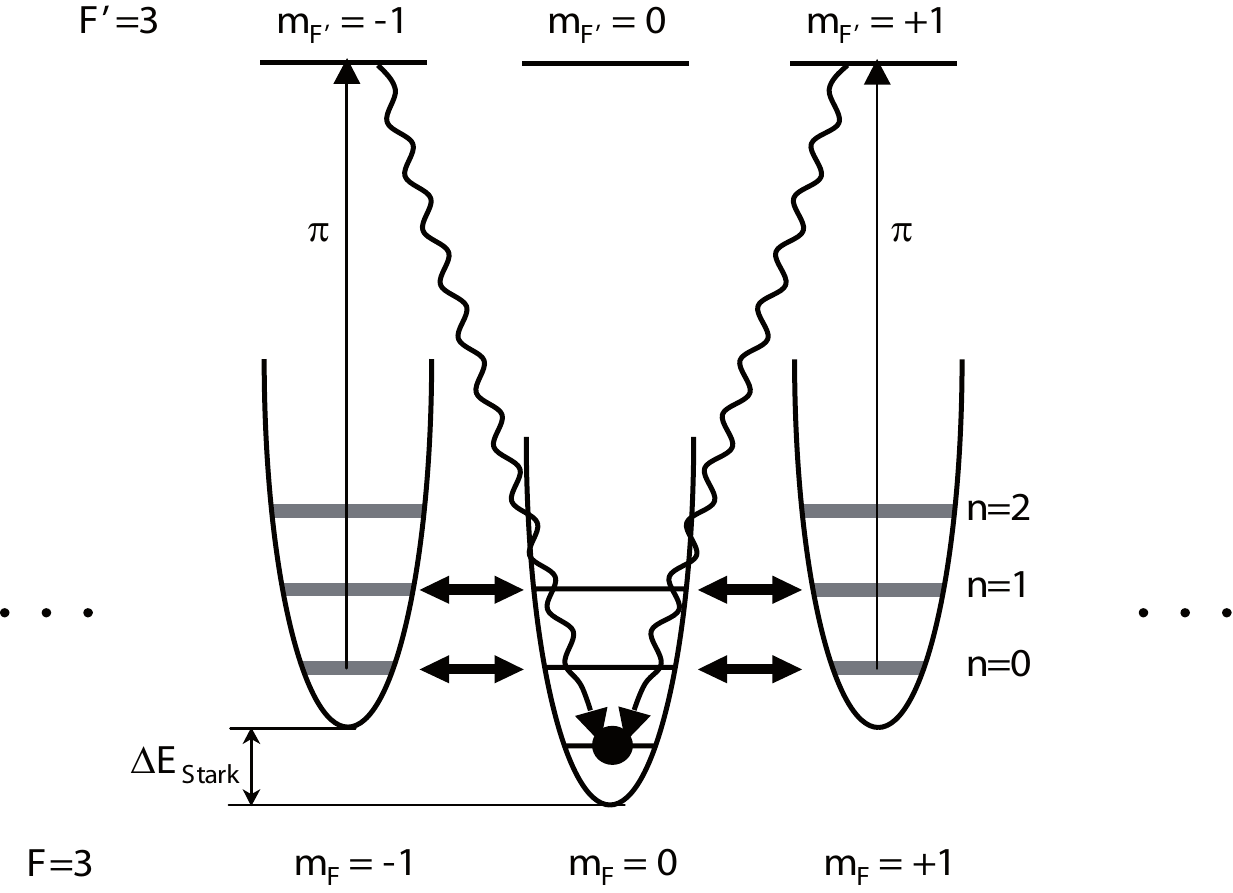}
\caption{Principle of Stark-shift-degenerate Raman sideband cooling \cite{TaichenachevPaper}. Cold atoms are trapped in the potential wells of an optical lattice, their motion is quantized, and $n$ is the vibrational quantum number. Ground state Zeeman sub-levels with $m\ne 0$ are light-shifted by the blue detuned $\pi$-pumping laser. When degeneracy is reached between $\left|F=3,m=0,n\right\rangle$ and $\left|F=3,m=\pm 1,n-1\right\rangle$, degenerate Raman sideband transitions can take place (black horizontal arrows), followed by $\pi$-pumping which closes the cooling cycle. Every cooling cycle removes one vibrational quantum until the atoms reach $\left|F=3,m=0,n=0\right\rangle$. See Ref.~\cite{TaichenachevPaper} for more details.}
\label{fig:Fig7}
\end{figure}
%

% Discuter le role de la frequence du laser pompe. 
% On peut refroidir plus efficacement en desaccordant un peu ce laser. 
% Tentative d'explication.

As shown in Fig.~\ref{fig:Fig5}, detuning the Zeeman pump laser on the blue side of the $3\rightarrow 3'$ transition can lead to a notable improvement of state preparation, which manifests itself as an increase of the $\left|F=3,m=0\right\rangle$ flux. As mentioned in section~\ref{section4}, the asymmetry observed around the $3\rightarrow 3'$ transition cannot be explained by optical pumping mechanisms only. Here the light shift produced by the pump laser plays an important role and our interpretation is that we observe the first signs of Stark-shift-degenerate Raman sideband cooling, as proposed in Ref.~\cite{TaichenachevPaper}. 

This cooling mechanism, which is briefly recalled in Fig.\ref{fig:Fig7}, is crucially dependent on the light-shift induced by the pump laser. Indeed, it produces cooling cycles for a positive light shift, and conversely heating cycles for a negative light shift. As a consequence, it will induce an asymmetric variation of state preparation efficiency around the $3\rightarrow 3'$ transition as observed in Fig.\ref{fig:Fig5}. 

In order to check the relevance of this explanation, we calculated the depth of the potential wells created by the optical lattice laser in our experimental conditions. We used the procedure developped in Ref.~\cite{DiDomenicoSidebandPaper} to obtain the potential depth $\Delta U\approx 5 E_r$ where $E_r=\hbar^2 k^2/(2M)$ is the recoil energy ($E_r\approx h\times 2$~kHz). Taking into account the anharmonicity of the potential wells, there are $3$ bound states and the vibrational frequency is $\omega_v\approx 2\pi\times 4$~kHz. In order to light shift the $m=\pm 1$ Zeeman sub-levels by $\hbar\omega_v$ we calculated that $15$~$\mu$W of pump power are necessary. This is close to the experimental values, which are between $3$~$\mu$W and $30$~$\mu$W, and therefore the mechanism of Stark-shift-degenerate Raman sideband cooling should be effective in our experiment. Here, one should keep in mind that these numbers represent average values, but the experimental situation is more complex since all the laser beams are gaussian with an intensity which varies by a factor $10$ from the center to the edge.

\begin{figure}[tbp]
\includegraphics[width=\grscale\textwidth]{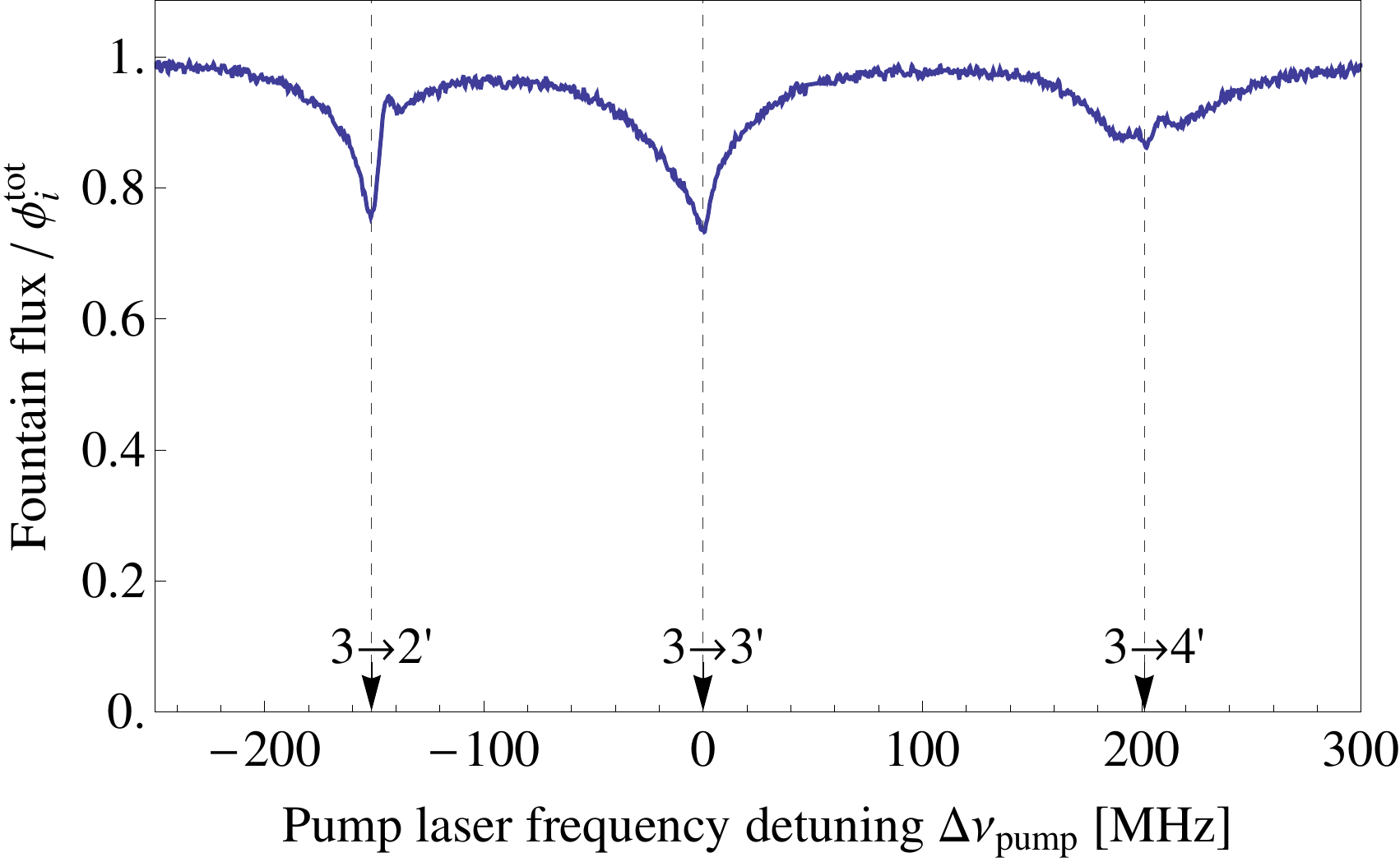}
\caption{Measurement of the total fountain flux as a function of the Zeeman pump laser frequency. $\Delta\nu_\mathrm{pump}$ is the pump laser frequency detuning from the $F=3\rightarrow F'=3$ transition of cesium D2 line. The pump laser power is $3$~$\mu$W. The vertical axis is normalized to the total flux obtained without state preparation $\phi^\mathrm{tot}_{i}$. The horizontal axis is calibrated using a saturated absorption signal obtained from a small fraction of the pump laser.}
\label{fig:Fig8}
\end{figure}

The Stark-shift-degenerate Raman sideband cooling mechanism not only prepares the atoms in $m=0$ but also cools them. Therefore, it should also have an effect on the total flux which gives an indirect measurement of the atomic beam transverse temperature. Indeed, we measured the total flux as a function of the pump laser frequency and the results are shown in Fig.~\ref{fig:Fig8}. One can see that the variation of the total flux is asymmetric around the $3\rightarrow 2'$ and $3\rightarrow 3'$ transitions, the cooling efficiency being higher on the blue side of those transitions as expected according to sideband cooling. This observation deserves two comments. Firstly, Stark-shift-degenerate Raman sideband cooling is also possible on the $3\rightarrow 2'$ transition where it accumulates the atoms in $m=\pm 3$ and thus depletes $m=0$ as shown in Fig.~\ref{fig:Fig5}. Secondly, the asymmetry in cooling efficiency is much more visible on the $3\rightarrow 2'$ transition which can be attributed to the fact that this transition is closed, i.e. the atoms are rarely pumped into the other ground state $F=4$ where Sisyphus cooling takes place. This is yet another evidence that a cooling mechanism is active on the $F=3$ ground state and that this mechanism involves light shifts.

To conclude this discussion, we remark that Stark-shift-degenerate Raman sideband cooling mechanism explains all the experimental features observed by varying the frequency of the Zeeman pump laser in our state preparation experiment. Nevertheless, present limitations of our experimental setup do not allow us to conclude with certainty. Further studies are necessary in order to confirm this interpretation.

%%%%%%%%%%%%%%%%%%%%%%%%%%%%%%%%%%%%%%%%%%%%%%%%%%%%%%%%%%%%%%%%%%%%%%%%%%%%%
\section{Conclusion}
\label{section6}
%%%%%%%%%%%%%%%%%%%%%%%%%%%%%%%%%%%%%%%%%%%%%%%%%%%%%%%%%%%%%%%%%%%%%%%%%%%%%

In this work, we demonstrate that quantum state preparation can be combined with laser cooling to prepare a continuous atomic fountain of cold atoms. More precisely, we use two-laser optical pumping to prepare cold cesium atoms in the same ground state $\left|F=3,m=0\right\rangle$. A first laser, in a folded optical lattice configuration, couples to the $F=4$ ground state and transfers the atoms in $F=3$, while a second $\pi$-polarized laser, the so-called Zeeman pump, excites the $F=3\rightarrow F'=3$ transition of the D2 line to pump the atoms toward $m=0$. When both lasers are on resonance, we observe a notable heating of the cold atomic beam produced by the optical pumping cycles. On the other hand, we demonstrate that it is possible to combine state preparation with Sisyphus cooling by detuning the optical lattice laser frequency, and thereby to remove most of the heat produced by optical pumping. 

Using this technique, we were able to prepare 56.6\% of the atoms in $\left|F=3,m=0\right\rangle$, limited by technical imperfections, without degrading the total fountain flux. Moreover the atomic orientation (asymetry of the population distribution among Zeeman sub-levels) of the prepared beam has been reduced by a factor twelve which is an advantage for its use in a primary frequency standard. Furthermore, we showed that state preparation improves the signal-to-noise ratio of the $\left|F=3,m=0\right\rangle$ fountain flux by a factor $\sqrt{3}$ which corresponds to a three-fold improvement of the shot-noise limited equivalent flux. At high flux, this signal-to-noise ratio is presently limited by some fluctuation in the beam intensity at the beam preparation stage. 

Finally, we observed an improvement of both laser cooling and state preparation efficiency when the Zeeman pump laser frequency is detuned on the blue side of the $F=3\rightarrow F'=3$ transition. In contrast, the efficiency decreases on the red side of the same transition, which reveals the role of the light-shift produced by the Zeeman pump laser. Similar observations where made around the $F=3\rightarrow F'=2$ transition, even though the final Zeeman states are $m=\pm3$ instead of $m=0$. We attribute this improvement of both laser cooling and state preparation efficiency to the first observation of Stark-shift-degenerate Raman sideband cooling.

\begin{acknowledgments}
This work was supported by the Swiss National Science Foundation (grant 200020-121987/1 and Euroquasar project 120418), the Swiss Federal Office of Metrology (METAS), and the University of Neuch\^atel. We acknowledge helpful comments from the referee.
\end{acknowledgments}

%%%%%%%%%%%%%%%%%%%%%%%%%%%%%%%%%%%%%%%%%%%%%%%%%%%%%%%%%%%

%%%%%%%%%%%%%%%%%%%%%%%%%%%%%%%%%%%%%%%%%%%%%%%%%%%%%%%%%%%
\end{document}